\documentclass[aps, prd, amsmath, floats, floatfix, twocolumn, nofootinbib, superscriptaddress, showpacs]{revtex4-1}
\usepackage{graphicx}
\usepackage{color}
\usepackage{latexsym}

\newcommand{\be}{\begin{eqnarray}}
\newcommand{\ee}{\end{eqnarray}}

\begin{document}

\title{Quasi-periodic oscillations as a tool for testing the Kerr metric: A comparison with gravitational waves and iron line}

\author{Cosimo Bambi}
\email{bambi@fudan.edu.cn}
\affiliation{Center for Field Theory and Particle Physics and Department of Physics, Fudan University, 200433 Shanghai, China}
\affiliation{Theoretical Astrophysics, Eberhard-Karls Universit\"at T\"ubingen, 72076 T\"ubingen, Germany}

\author{Sourabh Nampalliwar}
\email{sourabh@fudan.edu.cn}
\affiliation{Center for Field Theory and Particle Physics and Department of Physics, Fudan University, 200433 Shanghai, China}

\date{\today}

\begin{abstract}
QPOs are a common feature in the X-ray power density spectrum of black hole binaries and a potentially powerful tool to probe the spacetime geometry around these objects. Here we discuss their constraining power to test the Kerr black hole hypothesis within the relativistic precession model. We compare our results with the constraints that can be obtained from gravitational waves and iron line. We find that QPOs may provide very precise measurements, but they are strongly affected by parameter degeneracy, and it is difficult to test the Kerr metric with this approach in the absence of independent observations to constrain the mass or the spin.
\end{abstract}


\maketitle


\section{Introduction}

Up to now general relativity has been mainly tested by experiments in the Solar System and by observations of binary pulsars~\cite{will}. Constraints on possible deviations from the predictions of Einstein's theory of gravity in weak gravitational fields are today quite stringent. Tests of general relativity in the strong gravity regime are the new frontier.

The best laboratory for testing strong gravity is the spacetime around astrophysical black holes. According to general relativity, the spacetime around these objects is well described by the Kerr solution. Initial deviations from the Kerr metric are quickly radiated away by the emission of gravitational waves~\cite{kerr1}. The equilibrium electric charge is completely negligible for macroscopic objects~\cite{kerr2}. The presence of an accretion disk is irrelevant, and its impact on the background metric can be safely ignored~\cite{kerr3}.

There are two possible approaches to test the nature of black holes. We can study the properties of the electromagnetic radiation emitted by the accreting gas close to these objects~\cite{rev1,rev2,rev3}. Today the two leading techniques are the study of the thermal spectrum of thin disks~\cite{cfm1,cfm2,cfm3} and the analysis of the reflected spectrum (iron line)~\cite{iron1,iron2,iron3,iron4}, but new methods will be available in the future. The second approach is the study of the gravitational waves emitted from black holes~\cite{rev4,rev5}. This was only at the level of speculation up to very recently, but it is now a reality~\cite{ligo}.

Each approach has its own advantages and disadvantages, and it is sometime difficult to compare their constraints. The study of the electromagnetic radiation can only test the background metric, because the properties of the radiation can be related to the geodesic motion of the gas in the accretion disk and of the photons from the point of emission to that of detection in the flat faraway region. In other words, the electromagnetic radiation cannot distinguish a Kerr black hole in general relativity from a Kerr black hole in another theory of gravity, because the geodesic motion is the same~\cite{gr1}. Moreover, the accretion process around a black hole is quite a complex phenomenon: accurate measurements are only possible if we have the correct astrophysical model and the systematics under control, which may be very challenging. On the contrary, gravitational waves can directly probe the Einstein equations~\cite{gr2}, with the disadvantage that it is necessary to consider a specific theoretical framework and it is more difficult to perform model-independent tests. Moreover, parameter degeneracy limit the capability of this approach.

The recent detection of gravitational waves from the coalescence of two stellar-mass black holes by LIGO has opened a new window to test general relativity~\cite{ligo-gr}. An extended analysis of the constraints from GW150914 on a number of gravity theories has been presented in~\cite{yunes}. A more model-independent analysis is reported in~\cite{kz}, where the authors consider the quasi-normal modes of a scalar field on a deformed non-Kerr metric. Here the idea is that often -- but not always, depending on the specific gravity theory -- the frequency of the quasi-normal modes of a scalar field (which only depends on the background metric) are not very different from those of the gravitational waves (which instead can only be derived by the field equations of the gravity theory). The finding of Ref.~\cite{kz} is that the observation of GW150914 cannot rule out large deviations from the Kerr metric.

In Ref.~\cite{ale}, one of us has discussed the constraining power of the iron line method by employing the same metric as in Ref.~\cite{kz}, in order to compare the gravitational wave and the iron line approaches. While both the studies in~\cite{kz} and \cite{ale} are only preliminary analyses to get an idea of the potentialities of the two techniques, one can arrive at some interesting conclusions. The iron line method can potentially be quite competitive and provide stringent constraints. The reason is that one has to fit the whole shape of the iron line (actually the whole reflected spectrum, but most of the information on the strong gravity field is encoded in the iron line), while in the case of the gravitational waves one has just two numbers (in the case of GW150914) associated to the frequency of the observed quasi-normal mode. The weak point of the iron line is the astrophysical model, and there is not a common consensus that the iron line can really be used to get precise measurements of the spacetime metric.

The aim of this work is to further investigate the constraining capability of different techniques. We consider the quasi-periodic oscillations (QPOs) of the stellar-mass black hole in GRO~J1655-40 and we constrain possible deviations from the Kerr metric by employing the same metric as in~\cite{kz,ale}. Interestingly, current constraints would be at least comparable, but maybe even better, than the constraints from gravitational waves and iron line. However, there is a strong parameter degeneracy. Even if we assume to have much better measurements in the future, it is difficult to break the parameter degeneracy without independent measurements of the mass or of the spin.

\section{Testing the Kerr metric with QPOs}

QPOs are a common feature in the X-ray power density spectrum of black hole binaries~\cite{rem06,tbel}. There are several types of QPOs. Low-frequency QPOs are in the range 0.1-30~Hz and are divided in type-A, type-B, and type-C according to their properties. High-frequencies QPOs are in the range $\sim$100-500~Hz and some sources show an upper and a lower high-frequency QPO with the ratio 3:2. At the moment, there is no common consensus on the mechanism responsible for these QPOs. However, recent studies seem to support the relativistic precession model~\cite{motta1,motta2}, which associates the frequencies of some QPOs to the three fundamental frequencies of a test-particle in the background metric (orbital frequency, radial epicyclic frequency, vertical epicyclic frequency). The possibility of using QPOs to test the Kerr metric has been already investigated, see e.g.~\cite{qpo1,qpo2,qpo3,qpo4}.

Let us consider a generic stationary, axisymmetric, and asymptotically flat spacetime. We write the line element as
\be
ds^2 = g_{tt} dt^2 + g_{rr}dr^2 + g_{\theta\theta} d\theta^2 
+ 2g_{t\phi}dt d\phi + g_{\phi\phi}d\phi^2 \, . \nonumber
\ee
The metric coefficients are independent of the $t$ and $\phi$ coordinates, leading to the existence of the conserved specific energy at infinity, $E$, and the conserved $z$-component of the specific angular momentum at infinity, $L_z$. The $t$- and $\phi$-component of the 4-velocity of a test-particle can thus be written as
\be
\dot{t} = \frac{E g_{\phi\phi} + L_z g_{t\phi}}{
g_{t\phi}^2 - g_{tt} g_{\phi\phi}} \, , \quad 
\dot{\phi} = - \frac{E g_{t\phi} + L_z g_{tt}}{
g_{t\phi}^2 - g_{tt} g_{\phi\phi}} \, .
\ee
From the conservation of the rest-mass, $g_{\mu\nu}\dot{x}^\mu \dot{x}^\nu = -1$, we have
\be
g_{rr}\dot{r}^2 + g_{\theta\theta}\dot{\theta}^2
= V_{\rm eff}(r,\theta,E,L_z) \, ,
\ee
where the effective potential $V_{\rm eff}$ is
\be
V_{\rm eff} = \frac{E^2 g_{\phi\phi} + 2 E L_z g_{t\phi} + L^2_z 
g_{tt}}{g_{t\phi}^2 - g_{tt} g_{\phi\phi}} - 1  \, .
\ee

Circular orbits in the equatorial plane have $\dot{r} = \dot{\theta} = \ddot{r} = 0$. We write the geodesic equations as
\be
\frac{d}{d\lambda} \left( g_{\mu\nu} \dot{x}^\nu \right)
= \frac{1}{2} \left( \partial_\mu g_{\nu\rho} \right) \dot{x}^\nu \dot{x}^\rho \, ,
\ee
and we consider the radial component (namely $\mu = r$)
\be\label{eq-geo}
\left( \partial_r g_{tt} \right) \dot{t}^2 
+ 2 \left( \partial_r g_{t\phi} \right) \dot{t} \dot{\phi}
+ \left( \partial_r g_{\phi\phi} \right) \dot{\phi}^2 = 0 \, .
\ee
From Eq.~(\ref{eq-geo}) we obtain the orbital angular velocity $\Omega_\phi = \dot{\phi}/\dot{t}$
\be
\Omega_\phi = \frac{- \partial_r g_{t\phi} 
\pm \sqrt{\left(\partial_r g_{t\phi}\right)^2 
- \left(\partial_r g_{tt}\right) \left(\partial_r 
g_{\phi\phi}\right)}}{\partial_r g_{\phi\phi}} \, ,
\ee
where the sign is $+$ ($-$) for corotating (counterrotating) orbits. The orbital frequency is thus $\nu_\phi= \Omega_\phi/2\pi$

From $g_{\mu\nu}\dot{x}^\mu \dot{x}^\nu = -1$ with $\dot{r} = \dot{\theta} = 0$ we have
\be
\dot{t} = \frac{1}{\sqrt{-g_{tt} - 2g_{t\phi}\Omega_\phi - g_{\phi\phi}\Omega^2_\phi}} \, .
\ee
Since $-E = g_{tt} \dot{t} + g_{t\phi} \dot{\phi}$ and $L_z = g_{t\phi} \dot{t} + g_{\phi\phi} \dot{\phi}$, we find
\be
E &=& - \frac{g_{tt} + g_{t\phi}\Omega_\phi}{
\sqrt{-g_{tt} - 2g_{t\phi}\Omega_\phi - g_{\phi\phi}\Omega^2_\phi}} \, , \\
L_z &=& \frac{g_{t\phi} + g_{\phi\phi}\Omega_\phi}{
\sqrt{-g_{tt} - 2g_{t\phi}\Omega_\phi - g_{\phi\phi}\Omega^2_\phi}} \, .
\ee

The radial and vertical epicyclic frequencies can be obtained by studying small perturbations around circular equatorial orbits. If $\delta_r$ and $\delta_\theta$ are the small displacements around the mean orbit (i.e. $r = r_0 + \delta_r$ and $\theta = \pi/2 + \delta_\theta$), they are governed by the following differential equations
\be\label{eq-o1}
\frac{d^2 \delta_r}{dt^2} + \Omega_r^2 \delta_r = 0 \, , \quad
\frac{d^2 \delta_\theta}{dt^2} + \Omega_\theta^2 \delta_\theta = 0 \, ,
\label{eq-o2}
\ee
where
\be\label{eq-or}
\Omega^2_r = - \frac{1}{2 g_{rr} \dot{t}^2} 
\frac{\partial^2 V_{\rm eff}}{\partial r^2} \, , \quad
\Omega^2_\theta = - \frac{1}{2 g_{\theta\theta} \dot{t}^2} 
\frac{\partial^2 V_{\rm eff}}{\partial \theta^2} \, .
\label{eq-ot}
\ee
The radial epicyclic frequency is $\nu_r = \Omega_r/2\pi$. The vertical epicyclic frequency is $\nu_\theta = \Omega_\theta/2\pi$.

The periastron precession frequency $\nu_{\rm p}$ and the nodal precession frequency $\nu_{\rm n}$ can be obtained as
\be
\nu_{\rm p} = \nu_\phi - \nu_r \, , \quad
\nu_{\rm n} = \nu_\phi - \nu_\theta \, .
\ee

In the Kerr metric, all these frequencies only depend on three parameters: the black hole mass $M$, the spin parameter $a_*$, and the radius of the orbit $r$. According to Ref.~\cite{motta1} (see also Ref.~\cite{motta2}), the upper high-frequency QPO $\nu_{\rm U}$ would correspond to the orbital frequency $\nu_\phi$, the lower high-frequency QPO $\nu_{\rm L}$ would correspond to the periastron precession frequency $\nu_{\rm p}$, and the low-frequency type-C QPO $\nu_{\rm C}$ would correspond to the nodal precession frequency $\nu_{\rm n}$, namely
\be
\nu_{\rm U} = \nu_\phi \, , \quad
\nu_{\rm L} = \nu_{\rm p} \, , \quad
\nu_{\rm C} = \nu_{\rm n} \, .
\ee

In the case of the black hole binary GRO~J1655-40, there is an observation in which one detects the three QPOs above at the same time. Assuming the three frequencies correspond to the same fluid oscillation and are therefore produced at the same radial coordinate, one can solve the system of three equations (the expression of $\nu_\phi$, $\nu_{\rm p}$, and $\nu_{\rm n}$ in the Kerr metric) to infer the three unknown parameters ($M$, $a_*$, $r$). Since the QPO frequencies can be measured with a precision of order 1\%, one can determine $M$ and $a_*$ with a precision of $\sim$1\%~\cite{motta1}. Such a precision in the spin measurement is well above that from the iron line and gravitational waves. Moreover, if the model is correct, the approach is not affected by the astrophysical complications present for the iron line method.

\section{Constraints on deformations}

To compare the constraining power of the QPO approach with those of the gravitational waves in Ref.~\cite{kz} and of the iron line in Ref.~\cite{ale}, we use the same test-metric. The line element reads~\cite{kz}
\be\label{eq-m}
ds^2 &=& - \frac{N^2(r,\theta) - W^2(r,\theta) \sin^2\theta}{K^2(r,\theta)} dt^2
\nonumber\\ &&
- 2 W(r,\theta) \, r \sin^2\theta \, dt d\phi
+ K^2(r,\theta) \, r^2 \sin^2\theta \, d\phi^2
\nonumber\\ &&
+ \Sigma(r,\theta) \left[\frac{B^2(r,\theta)}{N^2(r,\theta)}dr^2 + r^2d\theta^2\right] \, ,
\ee
where
\be\label{eq-m2}
N^2(r,\theta) &=& \frac{r^2 - 2Mr + a^2}{r^2} - \frac{\eta}{r^3} \, , \nonumber\\
B^2(r,\theta) &=& 1 \, , \nonumber\\
\Sigma(r,\theta) &=& \frac{r^2 + a^2 \cos^2\theta}{r^2} \, , \nonumber\\
K^2(r,\theta) &=& \frac{\left(r^2 + a^2\right)^2 - a^2 \sin^2\theta 
\left(r^2 - 2Mr + a^2\right)}{r^2 \left(r^2 + a^2 \cos^2\theta\right)}
\nonumber\\ &&
+ \frac{\eta a^2 \sin^2\theta}{r^3 \left(r^2 + a^2 \cos^2\theta\right)}\, , \nonumber\\
W(r,\theta) &=& \frac{2Ma}{r^2 + a^2 \cos^2\theta} 
+ \frac{\eta a}{r^2 \left(r^2 + a^2 \cos^2\theta\right)} \, ,
\ee
and $a = J/M$ is the rotation parameter. This metric is obtained by deforming the Kerr metric by adding a static deformation $\eta$ such that
\be
M \rightarrow M + \frac{\eta}{2r^2}.
\ee
It is convenient to rewrite $\eta$ as
\be
\eta = r_0 \left(r_0^2 - 2 M r_0 + a^2 \right) \, ,
\ee
where $r_0$ is the radial coordinate of the event horizon of the black hole metric in~(\ref{eq-m}). If we write
\be
r_0 = r_{\rm Kerr} + \delta r = M + \sqrt{M^2 - a^2} + \delta r \, ,
\ee
we can use $\delta r$ as the deformation parameter to quantify possible deviations from the Kerr spacetime. If $\delta r = 0$, $r_0$ reduces to the radial position of the event horizon of a Kerr black hole. In the general case, $\delta r$ measures the difference of the radial coordinate of the event horizon with respect to that of a Kerr black hole with the same mass and spin.

\begin{figure*}[t]
\vspace{0.5cm}
\begin{center}
\includegraphics[type=pdf,ext=.pdf,read=.pdf,width=8.5cm]{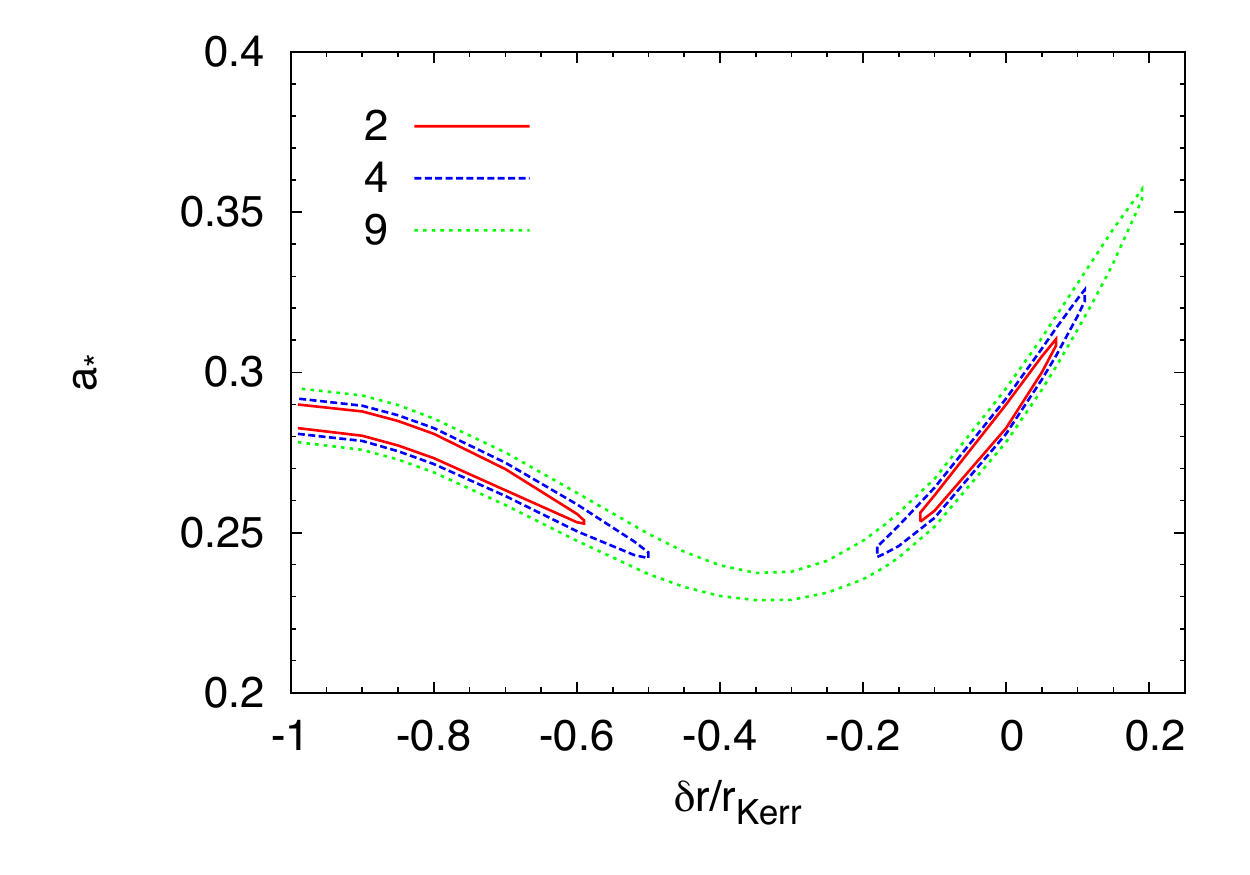}\\
\end{center}
\vspace{-0.5cm}
\caption{Constraints on the spin parameter $a_*$ and the deformation $\delta r/r_{\rm Kerr}$ for the black hole candidate in GRO~J1655-40 from current observations of QPOs within the relativistic precession model. The red-solid line, blue-dashed line, and green-dotted line represent, respectively, the contour levels $\Delta \chi^2 = 2$, 4, and 9. See the text for more details.\label{fig1}}
\vspace{0.5cm}
\begin{center}
\includegraphics[type=pdf,ext=.pdf,read=.pdf,width=8.5cm]{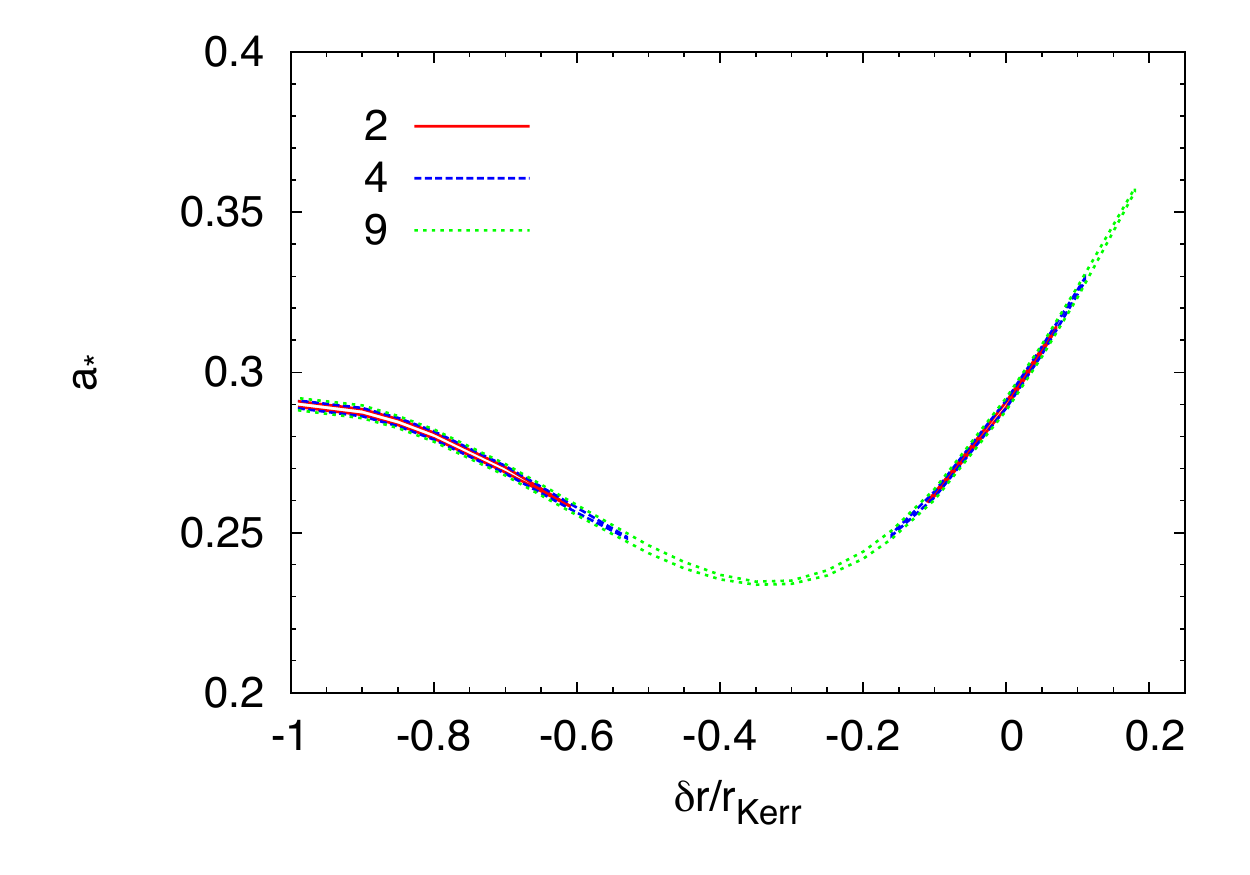}
\includegraphics[type=pdf,ext=.pdf,read=.pdf,width=8.5cm]{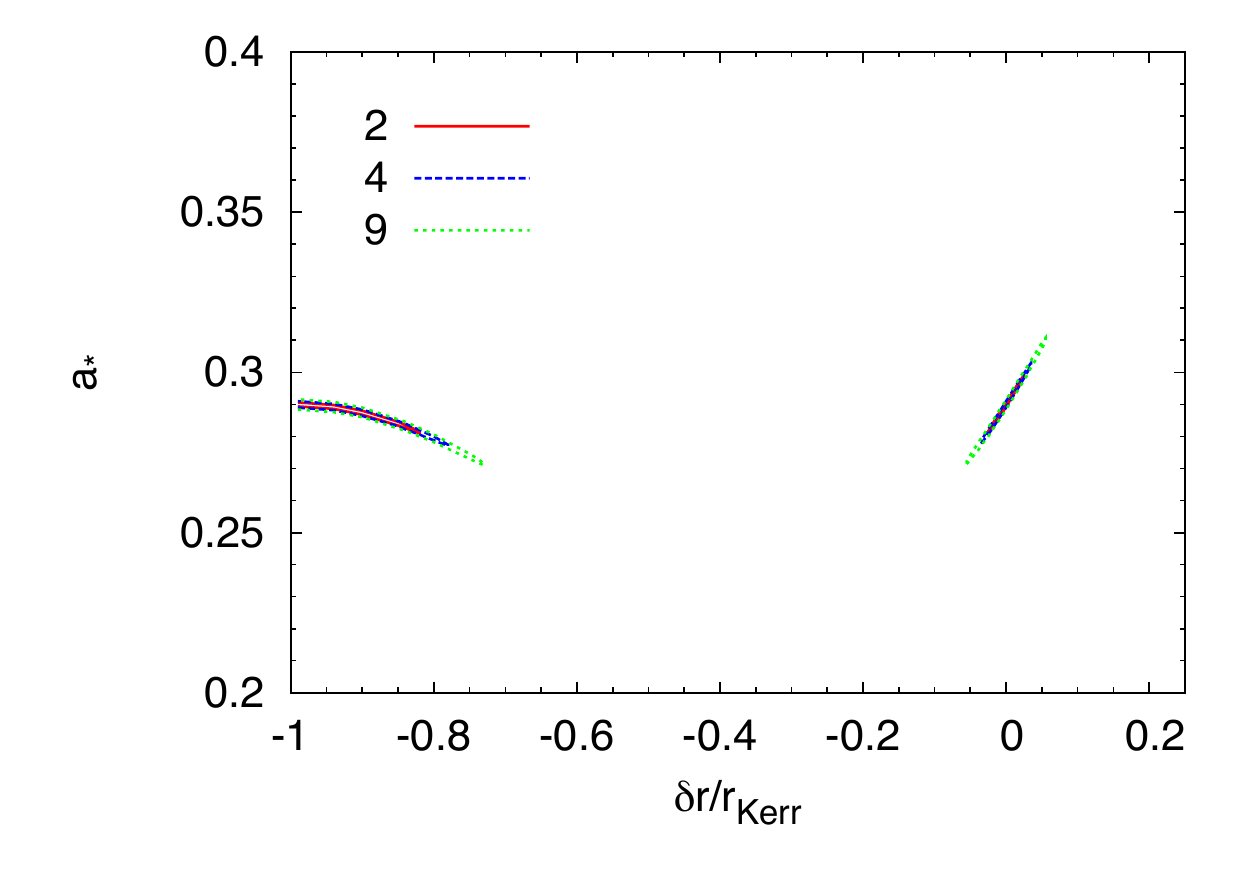}
\end{center}
\vspace{-0.5cm}
\caption{As in Fig.~\ref{fig1}, but assuming some hypothetical more precise future measurements. In the left panel, we consider the set of measurements in Eq.~(\ref{eq-set2}). The right panel shows the constraints from the set of measurements in Eq.~(\ref{eq-set3}). See the text for more details.\label{fig2}}
\end{figure*}

\subsection{Current observations}

As the first case, we want to find the constraints on $\delta r/r_{\rm Kerr}$ from the current observations of GRO~J1655-40. Presently, we have an observation in which we can measure all the three frequencies ($\nu_{\rm U}$, $\nu_{\rm L}$, and $\nu_{\rm C}$) and one in which we observe two frequencies ($\nu_{\rm U}$ and $\nu_{\rm C}$)~\cite{motta1}. Moreover, we have an independent dynamical measurement of the mass of the black hole~\cite{beer}. In summary, we have the following six measurements:  
\be\label{eq-set1}
&& \left( 441 \pm 2 , \, 298 \pm 4 , \, 17.3 \pm 0.1 \right) \, {\rm Hz} \, , \nonumber\\
&& \left( 451 \pm 5 , \, - , \, 18.3 \pm 0.1 \right) \, {\rm Hz} \, , \nonumber\\
&& M_{\rm dyn} = 5.4 \pm 0.3 \; M_\odot \, .
\ee
The free parameters are five: mass $M$, spin parameter $a_*$, radius of the observation with three frequencies, radius of the observation with two frequencies and deformation parameter. We perform a $\chi^2$ analysis as in Ref.~\cite{qpo2}. The contour levels of $\Delta\chi^2$ are shown in Fig.~\ref{fig1}. This plot can be compared with Fig.~1 in Ref.~\cite{kz} for the gravitational waves and with Figs.~1 and 2 in Ref.~\cite{ale} for the iron line.

\subsection{Better observations}

Let us assume that we can get better measurements of the frequencies than those available today and reported in~(\ref{eq-set1}). Assuming that GRO~J1655-40 is indeed a Kerr black hole with $a_* = 0.29$ and $M = 5.31$~$M_\odot$, we may have
\be\label{eq-set2}
&& \left( 441.4 \pm 1.0 , \, 298.0 \pm 1.0 , \, 17.59 \pm 0.05 \right) \, {\rm Hz} \, , \nonumber\\
&& \left( 451.0 \pm 1.0 , \, 313.1 \pm 1.0 , \, 18.36 \pm 0.05 \right) \, {\rm Hz} \, , \nonumber\\
&& M_{\rm dyn} = 5.4 \pm 0.3 \; M_\odot \, .
\ee
The frequencies are the same as in~(\ref{eq-set1}), but the associated uncertainty is smaller. They may be measurements with a future X-ray mission. The measurement of the mass is the same, because it is not obvious that we can have much better dynamical measurements in the future. We also note that (assuming the Kerr metric) the two sets of QPOs have a radial coordinate, respectively, of $r_1 = 5.67$~$M$, $r_2 = 5.59$~$M$. The constraints from this case are shown in the left panel in Fig.~\ref{fig2}. It is remarkable that the allowed region is now very thin. However, if we consider the allowed range of $\delta r/r_{\rm Kerr}$, it is almost the same as in Fig.~\ref{fig1}: thus the frequency measurements in~(\ref{eq-set2}), despite being much better than in (\ref{eq-set1}), do not help to break the degeneracy between the spin and possible deviations from the Kerr metric.

\subsection{More optimistic case}

The constraints found in the left panel in Fig.~\ref{fig2} from the set of measurements in~(\ref{eq-set2}) can be explained with the fact that in~(\ref{eq-set2}) we have essentially a precise measurement of the three fundamental frequencies at a specific radius, but not much more. Even if there are two sets of QPOs, they are generated almost at the same radius and therefore they provide almost the same information. It would be helpful to have instead the whole profile of the three frequencies. At the same time, the mass from dynamical measurement has quite a large uncertainty. A very precise measurement of $M_{\rm dyn}$ would be also helpful to break the parameter degeneracy, but it is unlike to get it at the necessary precision to do the job.

To show that this is indeed the right explanation, we consider a new set of measurements with the same precision as in~(\ref{eq-set2}), but now the difference between the two radii of the QPOs is larger. Our third set of measurements is
\be\label{eq-set3}
&& \left( 525.0 \pm 1.0 , \, 492.2 \pm 1.0 , \, 24.92 \pm 0.05 \right) \, {\rm Hz} \, , \nonumber\\
&& \left( 400.0 \pm 1.0 , \, 240.6 \pm 1.0 , \, 14.43 \pm 0.05 \right) \, {\rm Hz} \, , \nonumber\\
&& M_{\rm dyn} = 5.4 \pm 0.3 \; M_\odot \, ,
\ee
which corresponds to the frequencies expected from a Kerr black hole with spin parameter $a_* = 0.29$, mass $M = 5.31$~$M_\odot$, and where the radial coordinate is, respectively, $r_1 = 6.06$~$M$ and $r_2 = 5.04$~$M$. The constraints are reported in the right panel in Fig.~\ref{fig2}. Let us note, however, that a similar measurement may not be realistic. The upper and the lower high-frequency QPOs are always found in the ratio 3:2. This suggests some kind of resonance. It is possible that the signal is stronger when such a condition is satisfied, and weaker and more difficult to measure if the condition is not met. The ratio 3:2 is clearly possible only around some specific values of the radial coordinate, and this does not help to reconstruct the frequency profile.

\section{Concluding remarks}

Recent studies suggest that the QPOs in the X-ray power density spectrum of black hole binaries can be explained by the relativistic precession model. Since the frequencies associated with these QPOs can be measured with high precision and would be directly related to the fundamental frequency of the background metric (i.e. there are no complications related to the accretion flow), QPOs are potentially a powerful tool to test the nature of astrophysical black holes and general relativity in the strong gravity regime~\cite{qpo1,qpo2,qpo3,qpo4}.

In this letter, we have employed the metric proposed in~\cite{kz} and studied the constraints from present observations as well as from more precise, potentially achievable with upcoming instruments, measurements. Our results are summarized in Figs.~\ref{fig1} and \ref{fig2} and can be compared with the constraints from GW150914 reported in Fig.~1 in Ref.~\cite{kz} and those from the iron line in Figs.~1 and 2 in Ref.~\cite{ale}. Note that the constraints from gravitational waves and iron line in Refs.~\cite{kz,ale} are based on simplified analyses and should be considered as preliminary results, whereas the constraints shown here in Fig.~\ref{fig1} are the actual constraints from current observations of GRO~J1655-40. This is because the physics behind the QPOs -- assuming the relativistic model is correct -- is much easier than that involved in the other two approaches.

The constraining capability of these techniques clearly depends on the specific deviation from the Kerr metric under investigation, because different deformations alter different relativistic effects which, in turn, can have a stronger or a weaker impact on specific observable quantities. For example, Ref.~\cite{jjc-prd} shows that deformations in $g_{rr}$ cannot be constrained with a time-integrated measurement of the iron line, while stringent constraints can be obtained if we measure the temporal evolution of the iron line in response of a flare in the corona. In the case of the metric proposed in~\cite{kz}, deviations from the Kerr predictions mainly arise from the deformation of the metric coefficients $g_{tt}$, $g_{t\phi}$, and $g_{\phi\phi}$. Our conclusions generically hold for deformations of these metric coefficients, even if a different ansatz would quantitatively change the strength of the final constraints.

A correlated issue is whether the constraints from these different methods are independent or not. If we look at Figs.~\ref{fig1} and \ref{fig2} in the present paper, Fig.~1 in Ref.~\cite{kz}, and Figs.~1 and 2 in Ref.~\cite{ale}, the shapes of the allowed regions seem to be very similar. This is, again, because the metric coefficients $g_{tt}$, $g_{t\phi}$, and $g_{\phi\phi}$ produce quite strong observational effects in comparison to other metric coefficients by determining the characteristic orbits in the equatorial plane (innermost stable circular orbit, photon orbit, etc.). At first approximation, all these techniques are sensitive to these fundamental orbits. The shape of the constraints is similar because a typical deviation from Kerr shifts these orbits in the same direction and with a similar magnitude. As discussed in Ref.~\cite{cfm3}, the contour levels of these constraints essentially show which spacetimes have the same innermost stable circular orbits, photon orbits, etc. Only in the presence of high-quality data a technique like the iron line can be sensitive to smaller effects and breaks the degeneracy.

Current constraints from QPOs are quite competitive. This should not be a surprise considering that, assuming the Kerr metric, the spin measurement from QPOs has a precision of 1\%, $a_* = 0.290 \pm 0.003$~\cite{motta1}, while the measurement of the spin of the final black hole in GW150914 is at the level of 10\%, $a_* = 0.67^{+ 0.05}_{- 0.07}$~\cite{ligo}, and iron line spin measurements are at a similar level of the gravitational wave case.

All these approaches are affected by a strong correlation between the estimate of the spin and possible deviations from Kerr. In the case of the iron line, an accurate observation of the profile can break the parameter degeneracy and one can potentially get very strong constraints (Fig.~2 in \cite{ale}). This is because one has to fit the whole shape of the iron line, while in the case of QPOs and gravitational waves we only have a few numbers. In the latter cases, parameter degeneracy is quite natural. The disadvantage of the iron line method is instead related to the astrophysical model, and currently there is no common consensus on the possibility of using this technique to get very precise measurement of the metric.

In the case of QPOs, very precise measurements of the frequencies may not help much (see left panel in Fig.~\ref{fig2}). In the right panel in Fig.~\ref{fig2}, we have considered the possibility of QPOs generated at relatively different radii: this helps to constrain the metric, but it is not obvious that a similar detection is possible and the constraints would still be much weaker than that with the iron line from future X-ray missions. Furthermore, a very precise independent measurement of the mass and/or of the spin of the black hole would be helpful in constraining deviations from Kerr with the QPO approach.


\begin{acknowledgments}
This work was supported by the NSFC (grants 11305038 and U1531117) and the Thousand Young Talents Program. C.B. also acknowledges support from the Alexander von Humboldt Foundation.
\end{acknowledgments}


\end{document}